\documentclass[
 reprint,
superscriptaddress,
aps,
pra,
]{revtex4-2}

\usepackage{xfrac}
\usepackage{braket}
\usepackage{physics}
\usepackage{graphicx}
\usepackage{dcolumn}
\usepackage{bm}
\usepackage{subcaption}
\usepackage{graphicx}
\DeclareMathSizes{10}{10}{4}{4}

\usepackage[colorlinks=true, allcolors=blue]{hyperref}
\usepackage{floatrow}
\begin{document}

\title {\textit{De Novo} Design of Protein-Binding Peptides by Quantum Computing}
\author{Lars Meuser}\email{l.meuser@campus.unimib.it}
\affiliation{University of Milano-Bicocca, Piazza della Scienza 3, 20126 Milano, Italy.}
\affiliation{Sibylla Biotech S.p.A., Via Lillo del Duca 10,
20091 Bresso, Italy.}

\author{Alexandros Patsilinakos}
\affiliation{Sibylla Biotech S.p.A., Via Lillo del Duca 10,
20091 Bresso, Italy.}

\author{Pietro Faccioli}\email{pietro.faccioli@unimib.it}
\affiliation{University of Milano-Bicocca, Piazza della Scienza 3, 20126 Milano, Italy.}
\affiliation{INFN Sezione di Milano-Bicocca, Piazza della Scienza 3, 20126 Milano, Italy.}

\date{\today}

\begin{abstract}
\emph{In silico} \emph{de novo} design can drastically cut the costs and time of drug development. 
In particular, a key advantage of ``bottom-up'' physics-based approaches is their independence from training datasets, unlike generative models. However, they require the simultaneous exploration of chemical and conformational space. In this study, we address this formidable challenge leveraging quantum annealers. Focusing on peptide \emph{de novo} design, we introduce a multi-scale framework that integrates classical and quantum computing for atomically resolved predictions. We assess this scheme by designing binders for several protein targets. The D-Wave quantum annealer rapidly generates a chemically diverse set of binders with primary structures and binding poses that correlate well with experiments. These
results demonstrate that, even in their current early stages, quantum technologies can already empower physics-based drug design.

\end{abstract}

\maketitle

\section{Introduction}\label{intro}
Contemporary drug discovery research increasingly resorts to \emph{in silico} approaches to drastically reduce the time and costs of developing drug candidates.
For example, a customary approach in structure-based drug discovery consists of identifying hit candidates by performing virtual screening campaigns over libraries containing as many as $10^{10}$ compounds. The selected molecules must then be optimized to improve affinity, solubility, and deliverability and to reduce toxicity. This procedure is very time-consuming and expensive, involving several iterations of computer simulations and experiments. 

An alternative and potentially more efficient approach is one in which hit candidates are designed \textit{de novo}, taking into consideration the specific chemical environment provided by the binding pocket (for a recent review see, e.g.,~\cite{tang_recent_2024}). 
Algorithms for this purpose may assemble pre-selected molecular fragments~\cite{sommer_naominext_2019, metz_frag4lead_2021}, or update existing molecular structures with random mutations and cross-overs~\cite{spiegel_autogrow4_2020, yuan_ligbuilder_2020}. 
In recent years, several successful deep learning (DL) methods have been developed using a wide range of different neural network architectures (see, e.g.~\cite{AL_1_tilborg_traversing_2024, AL_2_korablyov_generative_2024, AL_3_dodds_sample_2024,CLM_1_iff_combining_2024,CLM_2_moret_leveraging_2023,diff_1_guan_3d_2023, diff_2_schneuing_structure-based_2024, reinf_1_gummesson_svensson_utilizing_2024, reinf_2_loeffler_reinvent_2024, geo_DL1_isert_structure-based_2023, VAE_1_wang_relation_2022, VAE_2_ragoza_generating_2022, MC_tree_1_ma_structure-based_2021,watson_novo_2023, pacesa_bindcraft_2024} and references therein).
At the same time, generative DL schemes still struggle to produce molecules with high affinity~\cite{mullard_when_2024} and synthesizabilty~\cite{gao_synthesizability_2020}.
Most importantly, most DL schemes require large target-specific databases and are biased toward generating molecules in the chemical neighborhood of the training set. 
While active learning schemes may help to tame this problem~\cite{AL_1_tilborg_traversing_2024, AL_2_korablyov_generative_2024, AL_3_dodds_sample_2024}, they come at a much higher computational cost. 

In contrast to DL-based schemes, ``bottom-up'' approaches based on modeling the statistical physics of the protein-ligand complex do not rely on training datasets. However, they require to explicitly account for the intra- and inter-molecular interactions.
Furthermore, finding optimal ligands poses a formidable optimization problem, as it involves simultaneously exploring chemical and conformational space.

The rapid development of quantum hardware raises the question of whether quantum computing could be integrated with physics-based models, helping to solve the underlying optimization problem of \emph{de novo} drug design.
To this end, a particularly attractive feature of quantum computers is that they enable the exploration of an exponentially large, combinatorially complex search space by exploiting quantum tunneling and superposition.
At the same time, several key tasks, including the mathematical representation of three-dimensional macromolecular structures and the implementation of classical force fields, are carried out more efficiently on classical computers. 

In this work, we develop a physics-based scheme to tackle the \textit{de novo} drug design problem, integrating classical and quantum computing to exploit their respective advantages.
As a first step in this direction, we focus on peptide-based drug design, optimizing both the sequence and binding pose to maximize binding affinity.
Peptide binders are an emerging class of drugs with applications in diverse therapeutic areas such as metabolic disorders and cancer~\cite{sharma_peptide-based_2023, barman_strategic_2023, wang_therapeutic_2022}.
Compared to small-molecule drugs, they are less toxic and more readily synthesizable, but have limited membrane permeability and are metabolized faster.
From a computational perspective, peptides have a simpler topology and a smaller number of building blocks than small molecules. Furthermore, empirical estimates of the affinity between the different amino acid types~\cite{MiyazawaJernigan1995} can be used to estimate the interaction energy at a coarse-grained level of resolution~\cite{kim_coarse_grained_2008}.
This enables us to employ a multi-scale approach, where the simultaneous exploration of chemical and conformational space is carried out at the coarse-grained level, reducing the number of binary variables required, and allowing us to solve the optimization problem on the D-Wave quantum annealer. The corresponding binding pose is then determined with a classical computer at full atomic resolution.

To validate our approach, we develop a statistical analysis to compare the predicted sequences and binding poses with those available in the experimental dataset.
We also compare the results from D-Wave's hybrid classical-quantum solver with those obtained by an industry-grade classical solver.

\section{Methods}\label{Section:theory}
\subsection{Statistical Physics Formulation of the Ligand Design Problem} 
From a statistical physics perspective, the general problem of identifying optimal ligands for given target protein $P$ can be formulated as 
\begin{eqnarray}
    \max_{\Gamma, \Sigma} \frac{\exp\left(-\frac{U(\Gamma, \Sigma;P)}{k_\text{B} T}\right)}{\sum_{P'}\sum_{\Gamma'}\exp\left(-\frac{U(\Gamma', \Sigma;P')}{k_\text{B} T}\right)}.
    \label{eq:sum_config_and_sequence}
\end{eqnarray}
In this expression, $U(\Gamma, \Sigma;P)$ denotes the interaction energy of the system consisting of a ligand of chemical composition $\Sigma$ in a configurational state $\Gamma$ and a protein $P$ in its native conformation.
The summation over all possible protein targets in the denominator ensures that the designed ligands \emph{selectively} maximize the affinity with the given target. We can equivalently reformulate the optimization problem~(\ref{eq:sum_config_and_sequence}) as
\begin{eqnarray}\label{exactCondition}
    \min_{\Gamma, \Sigma} \left( U(\Gamma, \Sigma; P)- G(\Sigma)\right),
\end{eqnarray}
where 
\begin{eqnarray}
    G(\Sigma) \equiv -k_\text{B} T\ln \sum_{P}\sum_ \Gamma \exp\left(-\frac{U(\Gamma, \Sigma;P)}{k_\text{B} T}\right) 
\end{eqnarray} 
is interpreted as the free energy associated to a given chemical structure $\Sigma$. 
Unfortunately, computing $G(\Sigma)$ is a formidable task because it involves accounting for all possible protein targets and, for each of them, summing the Boltzmann factors of all ligand configurational states. To reduce computational costs, we approximate its cumulant expansion truncated to the lowest order and obtain:
\begin{eqnarray}
    \min_{\Gamma, \Sigma}  \left( U(\Gamma, \Sigma;P)- \langle U(\Sigma)\rangle_0 \right), 
     \label{cumulant}
\end{eqnarray}
where $\langle U(\Sigma) \rangle_0 \equiv 
    \frac{1}{\mathcal{N}_S} \sum_P\sum_{\Gamma} U(\Sigma, \Gamma;P)$ and $\mathcal{N}_S \equiv   \sum_P\sum_{\Gamma} 1.$
According to the condition~(\ref{cumulant}), the optimal ligand is one that minimizes the binding energy with the given target $P$, relative to its \textit{average} interaction with proteins.
\subsection{Coarse-Grained Model}\label{Section:CG_model}
Let us now restrict our focus to the case in which the ligand is a peptide.
In the following, we develop a coarse-grained mathematical representation of the peptide's primary sequence $\Sigma$, the chain's conformational state $\Gamma$ and the interaction $U(\Gamma, \Sigma;P)$ that can be encoded on a collection of interacting two-level quantum systems (qubits).

\begin{figure*}
    \centering
    \includegraphics[trim=0 400 0 310, clip, width=\textwidth]{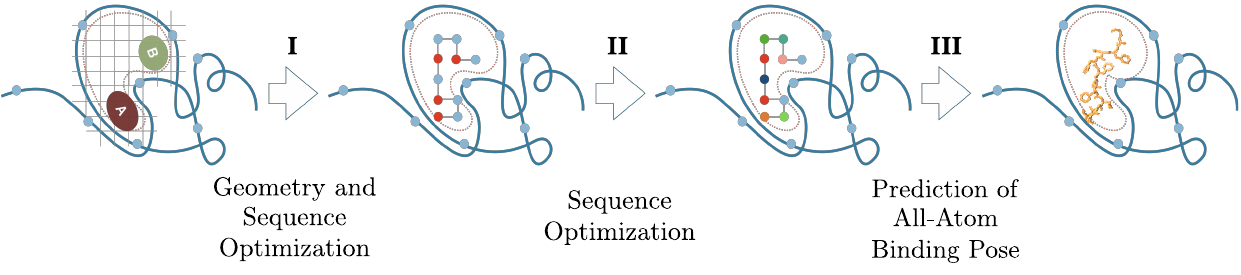}
    \caption{Schematic depiction of the workflow. I: A coarse-grained peptide connecting regions \textit{A} and \textit{B} is generated via (quantum) minimization of the problem Hamiltonian (Eq.~(\ref{eq:Hamiltonian_all_parts})). 
    II: The configuration is frozen and the sequence is optimized in higher resolution by a second (quantum) minimization step.
    III: A classical molecular docking simulation predicts the all-atom off-lattice representation of the previously generated sequence.}
    \label{fig:Schematic_algo}
\end{figure*}

{\bf Coarse-graining the chemical and conformational space:} We represent amino acids with single beads and group them into $D$ different chemical families. Furthermore, we discretize the peptide’s conformational space by introducing a square lattice that fills the pocket $P$ of the target protein (see leftmost panel in Fig.~\ref{fig:Schematic_algo}). The lattice spacing is set to match the length of the peptide bond (0.38\,nm), so that each configurational state of the peptide in the pocket can be identified with a self-avoiding path on the lattice.
In contrast, the protein's three-dimensional structure is represented using an off-lattice continuous model.

{\bf Coarse-graining the interaction:} To derive an expression for the interaction energy $U(\Sigma, \Gamma;P)$, we resort to the Miyazawa-Jernigan knowledge-based potential, first introduced in~\cite{MiyazawaJernigan1995}. Specifically, we follow the formulation proposed by Kim and Hummer~\cite{kim_coarse_grained_2008}, which includes Lennard-Jones (LJ) pairwise interactions between different amino acids in the protein and in the peptide.
To define such an interaction, we first introduce two $20\times20$ matrices of parameters, with entries indexed by $i,j\in \{1\,\ldots\,20 \}$
\begin{enumerate}
  \item An energy matrix $\hat \epsilon$ with entries:
\begin{align}\label{eq:transformed_MJ}
\varepsilon_{ij}=\lambda(e_{i j}-e_{0})
\end{align}
where $\lambda=0.159$ provides an overall scale, and $e_{i j}$ is the entry of the original Miyazawa-Jernigan matrix reported in Ref.~\cite{MiyazawaJernigan1995}, and $e_0 =-2.27 k_\text{B} T$ is an overall energy offset.
\item An interaction range matrix with entries: 
\begin{eqnarray}
\label{eq:radii}
\sigma_{ij} &=&\frac{\sigma_i + \sigma_j }{2},
\end{eqnarray}
where $\sigma_i$ denotes the
van der Waals (vdW) diameter of the amino acid of type
$i$ (the numerical values are reported in~\cite{kim_coarse_grained_2008}).
\end{enumerate}
The LJ interaction between an amino acid of type $i$ and one of type $j$ at a relative distance $r$ is given by:
\begin{eqnarray}
 u_{ij}(r) =
\begin{cases}
 4\lvert\varepsilon_{ij}\rvert\ \Big[ \big(\frac{\sigma_{ij}}{r}\big)^{12} -\big(\frac{\sigma_{ij}}{r}\big)^{6} \Big], & 
\varepsilon_{ij}<0\\
     4 \varepsilon_{ij} \Big[ \big(\frac{\sigma_{ij}}{r}\big)^{12} -\big(\frac{\sigma_{ij}}{r}\big)^{6} \Big] + 2\varepsilon_{ij},&   \epsilon_{ij}>0,\, r < r^0_{ij} \\
      -4\varepsilon_{ij} \ \Big[ \big(\frac{\sigma_{ij}}{r}\big)^{12} -\big(\frac{\sigma_{ij}}{r}\big)^{6} \Big],& \epsilon_{ij}>0, \, r \geq r^0_{ij}.
    \end{cases},\nonumber\\
    \label{eq:LJ_potential}
\end{eqnarray}
where $r^0_{ij} \equiv 2^{1/6}\sigma_{ij}$. We set a cut-off for all LJ interactions at 8.5\AA, a choice in line with the values adopted in binary contact maps based on $\text{C}_{\alpha}$ distances.

{\bf Reduced chemical alphabet:}
The model defined so far distinguishes between the 20 different types of naturally occurring amino acids. 
However, the chemical alphabet of amino acids is known to be redundant~\cite{li_reduction_2003} as multiple residues share similar physicochemical properties, 
such as, e.g. electric charge, polarity, and hydrophobicity. 
This redundancy can be exploited to develop an even coarser-grained approach in which amino acids are grouped into $D<20$ families. Theoretical studies have suggested that optimal grouping should include 5 to 10 families~\cite{chan_folding_1999, li_reduction_2003} to preserve most of the structural information.

Deriving this new representation amounts to mapping the $20 \times 20$ energy matrices $\hat{e}$ and ${\hat \sigma}$ onto $D\times D$ effective matrices $\hat {e}'$ and $\hat {\sigma}'$. 
The assignment of the 20 amino acids to the $D$ clusters can be done by finding the mapping $a(i) \in\{1, \ldots, D\}$ that minimizes the loss function
\begin{equation}
    L = \sum_{i,j=1}^{20}(e_{ij}-e'_{a(i)a(j)})^2.
\end{equation}
The entries of the corresponding effective (clustered) interaction matrix are defined by a mean over the elements of the clusters, i.e.
\begin{equation}
   e'_{ij}=\frac{1}{\mathcal{N}}\sum_{k,l=1}^{20}e_{kl} \, \delta_{a(k),i} \delta_{a(l),j}
\end{equation}
with $\mathcal{N}$ normalizing over the number of elements included in the two sets.
Similarly, the effective interaction range matrix is obtained by averaging over the elements of a cluster, i.e.: 
\begin{equation}
\sigma'_I = \frac{1}{\mathcal{N}}\sum_{k=1}^{20} \delta_{a(k),I}\sigma_{k}.
\end{equation}
We checked that this clustering procedure generates groups with similar physicochemical properties. For example, by choosing $D=2$ we obtain a bipartition of the amino acids that identifies hydrophobic residues.

{\bf Design optimization condition in the coarse-grained model:} To solve the design problem~(\ref{cumulant}), we need to evaluate the average interaction of the peptide $\Sigma$ with proteins, $\langle U(\Sigma)\rangle_0$.
To estimate this term using a manageable amount of qubit resources, we introduce a mean-field approximation: 
\begin{align}
    \langle U(\Sigma)\rangle_0 
    &\approx \mathcal{N}_c\sum_{n=1}^{l_\Sigma}\sum_{j=1}^{20}f_j\,\varepsilon_{i(n) j} 
    \label{eq:approx_sequence_FE}
\end{align}
In this equation, $l_\Sigma$ is the peptide length, $i(n)$ is the type of amino acid at position $n$ along the chain. $f_j$ is the relative frequency of the amino acid type $j$ on the surface of a typical protein, obtained from Ref.~\cite{miller_interior_1987} and $\mathcal{N}_c$ is the average number of contacts that a peptide residue forms with the amino acids on a typical protein surface (estimated in Section \ref{SM:AverageContacts} of the Supplementary Material (SM)).
We note that in Eq.~(\ref{eq:approx_sequence_FE}) the average interaction the amino acid sequence $\Sigma$ forms with typical protein surfaces is estimated by the interaction it would form on a fictitious average protein surface.

\subsection{Quantum Encoding of the Design Optimization Problem }\label{Section:quantum_encoding}
To be able to use the D-Wave quantum annealer to solve the peptide binder design problem, we encode condition~(\ref{cumulant}) as a Quadratic Unconstrained Binary Optimization (QUBO) problem. 
This requires mapping favorable peptide sequences and binding poses onto the low-energy states of a suitably defined quantum Hamiltonian, $\hat H$.
To this end, it is convenient to first establish a QUBO encoding based on classical binary variables (bits), and then to promote the formulation to the quantum level, replacing them with qubits. 

We introduce a collection of binary variables $q^{(k)}_i\in\{0,1\}$ at each grid point $i$, which are set to 1 if the site $i$ is occupied by a residue of type $k\in\{1, \ldots, D\}$.
Additional binary variables $q_{ij}$ denote the formation of a chemical bond between the residues at neighboring grid points $i$ and $j$.
Lastly, we shall resort to a set of ancillary variables $q^{(k)}_{ij}$ that are required to ensure that the Hamiltonian is, at most, quadratic in the binary variables.

The overall structure of our classical QUBO Hamiltonian consists of several terms: 
\begin{equation}
H=H_{\text{int}}+H_{\text{ext}}+H_{\text{anc}}+H_{\text{occ}}+H_{\text{path}}+H_{\text{chain}}
\label{eq:Hamiltonian_all_parts}. 
\end{equation}
The first two terms, $H_{\text{int}}$ and $H_{\text{ext}}$, represent the interactions of the peptide with itself and with the residues in the pocket, respectively. In particular, the latter is given by
\begin{equation}
    H_{\text{ext}} = \sum'_{i}\sum_{k=1}^D (E^{(k)}_{i}-E^{(k)}_0) q^{(k)}_{i},
\end{equation}
where $\sum'_{i}$ indicates the sum over grid points. $E^{(k)}_{i}$ is the (pre-computed) energy an isolated amino acid of type $k$ would experience at lattice site $i$ due to the interaction with the target protein's amino acids in the pocket.
$E^{(k)}_0$ accounts for condition (\ref{cumulant}), corresponding to the average interaction this isolated amino acid forms with a generic protein surface, and it is evaluated according to Eq.~(\ref{eq:approx_sequence_FE}).
$H_\textrm{int}$ is the Hamiltonian accounting for non-bonded \emph{intra-chain} interactions within the peptide and reads
\begin{align}
    H_{\text{int}} &= \sum'_{i,j
}\sum_{k, l=1}^D u_{kl}(r_{ij})\Big(q^{(k)}_i -q^{(k)}_{ij}\Big)\, q^{(l)}_j,
\label{eq:Interaction_Hamiltonian}
\end{align}
where $r_{ij}$ denotes the Euclidean distance between lattice site $i$ and $j$.
The ancillary variables $q^{(k)}_{ij}$ are defined on neighboring sites $i$ and $j$ only, and are otherwise set to 0. For neighboring sites they are set to 1 when the residue of type $k$ at site $i$ is involved in a chemical bond with a residue of any type at the neighboring site $j$, i.e., if $q_{ij}=q^{(k)}_i=1$.
This consistency condition is enforced by the Hamiltonian
\begin{align}\label{eq:H_anc}
    H_{\text{anc}} &= A \sum'_{\langle i,j\rangle}\sum_{k=1}^D \Big(3q^{(k)}_{ij} + q^{(k)}_iq_{ij} \notag \\ 
 &\quad -2q^{(k)}_iq^{(k)}_{ij} -2q_{ij}q^{(k)}_{ij}\Big), 
\end{align}
where $\sum'_{\langle i,j\rangle}$ represents the sum over neighboring lattice sites and $A$ is a positive constant that sets the overall energy penalty for violating the constraint. 
Note that, with this definition, the factor $\Big(q^{(k)}_i -q^{(k)}_{ij}\Big)$ excludes the interactions between covalently bonded amino acids.

The term $H_{\text{occ}}$ in Eq.~(\ref{eq:Hamiltonian_all_parts}) ensures that each grid point is occupied by at most one amino acid and reads
\begin{equation}
H_{\text{occ}}= A\sum'_{i}\sum^D_{k\neq l} q^{(k)}_{i}q^{(l)}_{i}.
\label{eq:H_occ}
\end{equation}

$H_{\textrm{path}}$ is defined to enforce the peptide's linear topology, i.e., to ensure that the internal residues along the chain are bonded to exactly two neighbors, while those at the terminals form a single bond:
\begin{eqnarray}
H_{\textrm{path}} &=& A \,(h_t+h_s+h_r),\\
\label{eq:path_hamiltonian}
    h_s &=& -\Big(1-\sum^D_{k=1} q^{(k)}_s \Big)^2 + \Big(\sum^D_{k=1}q^{(k)}_s-\sum'_{j\in \langle s,j\rangle} q_{sj}\Big)^2 \label{eq:H_s}\qquad\\
    h_t &=& -\Big(1-\sum^D_{k=1} q^{(k)}_t \Big)^2 + \Big(\sum^D_{k=1}q^{(k)}_t-\sum'_{j\in \langle t,j\rangle} q_{tj}\Big)^2 \label{eq:H_t}\qquad \\
    h_r &=& \sum'_{j\neq s,t} \Big(2\sum^D_{k=1}q^{(k)}_r-\sum'_{j\in \langle r,j\rangle} q_{rj}\Big)^2
    \label{eq:H_r}
\end{eqnarray}
The terms $\Big(1-\sum^D_{k=1} q^{(k)}_{s(t)} \Big)^2$ in Eqs.~(\ref{eq:H_s}) and~(\ref{eq:H_t}) specify the location of the chain endpoints, while $\Big(2\sum^D_{k=1}q^{(k)}_r-\sum'_{j\in \langle r,j\rangle} q_{rj}\Big)^2$ takes care of the chain's continuity requirement.
We note that, in the limiting case of just one chemical species, $H_\textrm{path}$ coincides with the QUBO Hamiltonian introduced in~\cite{krauss_solving_2020} to identify paths connecting two given nodes in a discrete network.

For the Hamiltonians~(\ref{eq:H_anc}),~(\ref{eq:H_occ}), and~(\ref{eq:path_hamiltonian}) to encode hard constraints, the energy scale $A$ must be large compared to all soft interactions.
Under this condition, low-energy configurations of the Hamiltonian~(\ref{eq:path_hamiltonian}) are those representing a linear chain that connects the lattice sites $s$ and $t$.

In addition to the chain, some of these states may also include topologically disconnected circularized peptides, which may be removed in post-processing. Alternatively, they can be suppressed by introducing additional penalty terms in $H$, specifically designed to penalize circular structures~\cite{micheletti_polymer_2021, panizza_statistical_2024}. Finally, spurious polymer rings are also suppressed when the total number of bonds (chain length $L_0$) is comparable to the shortest distance between the lattice sites $s$ and $t$. $L_0$ can be fixed by introducing an additional constraint
\begin{equation}
    H_{\text{length}}= w\Big(L_0-\sum'_{\langle i,j\rangle}q_{ij}\Big)^2
    \label{eq:quadratic_chain_penalty}.
\end{equation}
The factor $w$ can be tuned according to the desired tolerance on the length of the generated chains. To ensure a given fixed length, $w$ needs to be set comparable to $A$. In contrast, the choice $w \sim \frac{A}{L_0^2p^2}$ allows for generating chains with slightly different lengths, with relative fluctuations of the order of $p$ percent, i.e. $L \in [L_0(1-p), L_0(1+p)]$.

The advantage of the present formulation over the approach taken in Ref.~\cite{Danial_sampling_2022} is that it does not require fine-tuning of the weights and allows control over the magnitude of the fluctuations in the length of the generated linear peptides.
However, a disadvantage is the introduction of all-to-all connectivity between the bond variables, potentially making the optimization problem harder to tackle, both with classical and quantum computers.

In the SM we show that, for a cubic lattice of dimensions $L_x, L_y, L_z$, the total number of binary variables required is
\begin{equation}
    N_\text{qubits}\sim L_xL_yL_z(4D+3).
    \label{eq:qubit_estimate}
\end{equation}

\textbf{Classical and Quantum Algorithms for QUBO:}
The QUBO problem defined above can be solved using classical and quantum hardware.
Classical optimization schemes may combine heuristic global search algorithms (such as simulated annealing) with local refinements. Other classical optimization methods, such as the branch-and-bound algorithm used in the Gurobi optimizer~\cite{gurobi}, provide a more systematic exploration of the search space.
Alternatively, QUBO problems may be tackled using quantum hardware, which capitalizes on quantum superposition to enhance the exploration of the configuration space.

To solve the peptide design problem, in this study, we resorted to both classical and quantum optimizers, comparing the results obtained using Gurobi (a state-of-the-art solver widely used in academic and industrial research) and the D-Wave quantum annealer.
In particular, to implement our QUBO problem on the D-Wave quantum annealer, it is convenient to recast the Hamiltonian $H$ in the form of a (classical) generalized Ising model. To this end, we apply the transformation $\sigma^z _l = 2 q_l - 1$, where the label $l$ runs over all binary variables entering the QUBO Hamiltonian. 
The resulting generalized Ising Hamiltonian contains both quadratic and linear terms, i.e., it takes the form $H_{\text{Ising}}= \sum_l h_l\sigma^z_{l}+\sum_{l>m}J_{lm}\sigma^z_{l} \sigma^z_{m}$. 
The classical Hamiltonian is then promoted to a quantum Ising Hamiltonian by replacing each spin variable with a Pauli-$z$ operator, $\sigma_l^z\rightarrow \hat \sigma_l^z$. The eigenstates of the $\hat \sigma_l^z$ operators are identified with the qubits of the quantum computer. 

In this quantum encoding, our peptide design problem is mapped onto finding the ground state of a generalized quantum Ising Hamiltonian $\hat H_{\text{Ising}}$. This task is conveniently tackled by resorting to the so-called adiabatic switching procedure~\cite{das_quantum_2005}:
The quantum computer's wave function is initialized in the ground state of a Hamiltonian that is easy to solve and does not commute with $\hat\sigma^z$, for example, $\hat{H}_{\textrm{in}}= -h \sum_l \hat \sigma^x_l$, 
where $h$ is an arbitrary real constant.
Then, the quantum annealer's wave function is evolved for a time $t_f$ according to the time-dependent Hamiltonian $\hat H(t) = a(t) \hat H_{\textrm{in}} + b(t) \hat H_\textrm{Ising}$. The so-called scheduling functions $a(t)$ and $b(t)$ are defined in such a way such that $\hat H(t)$ switches from $H_{\textrm{in}}$ to $\hat H_\text{Ising}$ over the time interval $t_f$, i.e., $a(0)\gg b(0)$ and $a(t_f)\ll b(t_f)$.
The adiabatic theorem ensures that if the sweeping process is performed sufficiently slowly compared to the minimal energy gap $\Delta E$ encountered (i.e. if $t_f \gg \frac{\hbar}{\Delta E}$), then the system remains in its instantaneous ground state. That is, measuring the qubits in the final quantum state yields a solution to the QUBO problem.

In our applications to realistic design problems, we resorted to the hybrid solver of the D-Wave quantum annealing machine, which combines quantum annealing with classical pre- and post-processing steps~\cite{Dwave_manual}.

\subsection{Peptide Design Algorithm}\label{subsec:workflow}
In summary, our peptide design algorithm operates according to the following multi-step procedure, which is schematically illustrated by Fig.~\ref{fig:Schematic_algo}:
\begin{enumerate}
\item Using the binary encoding described in Section~\ref{Section:quantum_encoding} and the coarse-grained model defined in Section~\ref{Section:CG_model}, we perform a simultaneous optimization of both the chain's primary sequence and its binding pose. 
To meet the limitations on the maximum number of qubits available on the existing quantum computing hardware, in this phase, we resort to the clustering algorithm defined in Section~\ref{Section:CG_model} to restrict the chemical alphabet to $D$ families, with $5\lesssim D\lesssim 10$. 
\item The location of the amino acids on the lattice obtained after the minimization in the previous step is held fixed, enabling more qubit resources to become available for a second, more refined optimization of the primary sequence that includes the full range of 20 amino acids. Fixing the conformation allows us to drop the ancillary qubits, as well as the terms $H_\text{anc}$, $H_\text{path}$, and $H_\text{chain}$.
\item The selected chain sequence $\Sigma$ is then passed to a state-of-the-art docking software which returns the off-lattice, atomistically detailed binding pose. In this work, we resorted to \textit{Autodock CrankPep} (ADCP) ~\cite{zhang_autodock_CP_2019}, a specialized version of the \textit{Autodock} software package designed for peptide docking.
\end{enumerate}

\section{Results and Discussion}\label{Section:Application}
In this section, we report several applications of our peptide design algorithm, and we provide a first assessment of its accuracy based on experimentally resolved protein-peptide complexes.
We also compare the performance of classical and quantum optimization in solving our QUBO problem.

Even though many experimentally resolved structures for protein-peptide complexes are available as PDB files, exploiting this information to assess the accuracy of our peptide design algorithm is not straightforward. Indeed, since the peptides' chemical space grows exponentially with the number of residues, it is extremely unlikely for any design algorithm to yield the sequences found in any of the available protein-peptide PDB entries. Likewise, the chains in the experimentally resolved complexes are not necessarily those with the largest binding affinity.

To overcome this problem and to meaningfully assess our algorithm, we devised two independent statistical analyses, which focus on the structure of the binding pose and on the peptides' primary structure, respectively.
\subsection{Structure-Based Validation}\label{subsec:Structure_Validation}
\begin{figure*}
    \centering
    \includegraphics[trim=70 10 0 23, clip, width=1.15\textwidth]{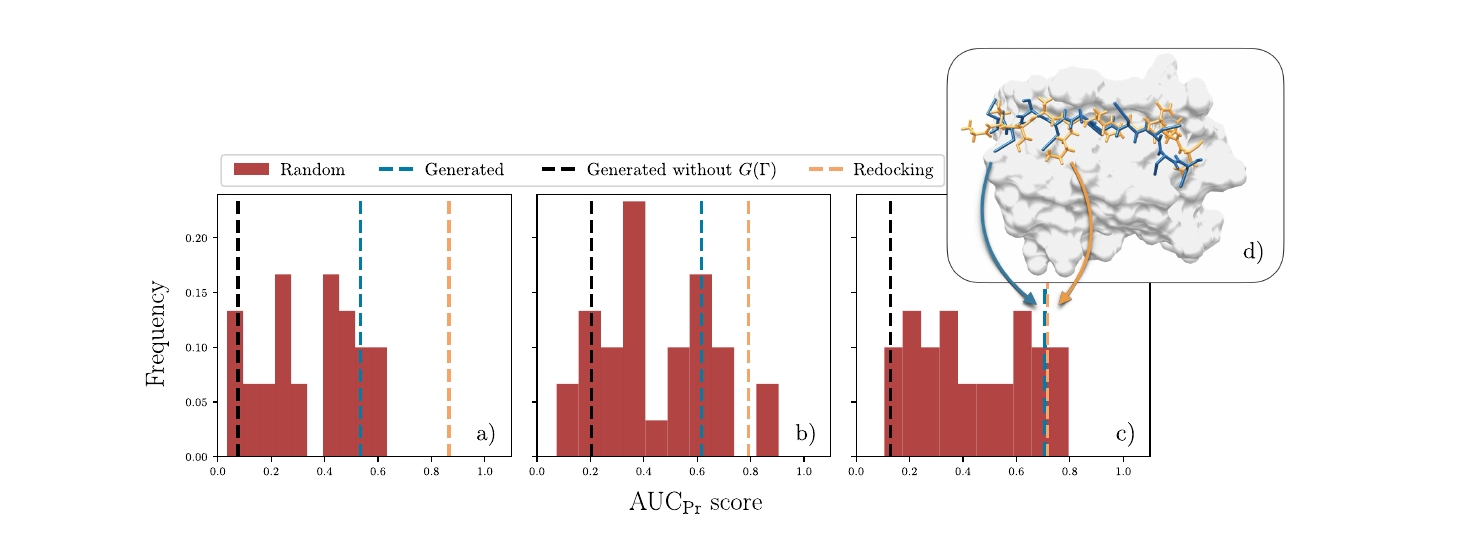}
    \caption{The area under the precision-recall curve for docking simulations of the PDB entries (a) 3BFW, (b) 4DS1 and (c) 3BRL.
    For each simulation, the scores of 30 random peptides (red) were compared to the previously removed peptide (orange), a generated peptide optimized without accounting for the sequence free energy (black), and a generated peptide optimized accounting for it (blue).
    A higher score indicates better binding to the pocket in the framework of the docking validation (see main text and Section~\ref{SM:docking_validation} of the SM). The redocking pose and the generated molecule's docking pose for 3BRL are shown in (d).}
    \label{fig:histogram_AUC_PR_all}
\end{figure*}

In principle, the quality of our algorithm may be assessed by checking if the binding free energies of the designed peptides are comparable to that of the peptide in the PDB structure and much larger than those of randomly generated peptides. In practice, this comparison would be flawed by the large systematic errors that affect ADCP's estimate of absolute binding free energies. 
However, ADCP performs well at structural predictions for a ligand's binding pose and in ranking alternative binding poses of the same ligand based on their estimated relative binding free energy~\cite{weng_comprehensive_2020}. 

These two features can be leveraged to devise a precision-recall analysis that assesses the quality of the designed peptides by comparing structural predictions rather than absolute binding free energies. 
To this end, we assume that the key interactions made by the peptide found in the PDB structure (referred to as the peptide's native contacts) are universal, i.e, common to all ligands that bind to the given pocket. In other words, good binders are assumed to form a high fraction of native contacts, $f_{\text{nat}}$.

To assess our algorithm, for each generated sequence, we used ADCP to produce a list of alternative predicted binding poses and checked if the poses with the largest values of $f_{\text{nat}}$ were ranked high in this list.
To quantify this test, we conducted a precision-recall analysis, marking poses with $f_{\text{nat}}>0.5$ as positive, a criterion also used in the Critical Assessment of Predicted Interactions (CAPRI)~\cite{collins_capri-q_2024}.
 
We considered three different protein-peptide complexes (PDB codes: 3BRL, 4DS1, 3BFW) taken from the LeadsPep dataset~\cite{hauser_leads-pep_2016}, with peptides containing 10 and 11 amino acids.
First, we removed the peptides from the PDB files and set up a lattice in the corresponding protein pocket, placing it in regions within 7.6 \AA \, of the $\text{C}_\alpha$ atoms in the experimental binding pose. 
Next, we chose the lattice sites $s$ and $t$, assigning them to be the grid points closest to the endpoints of the experimentally bound peptide. A more general and computationally expensive approach would be to compare the results obtained while also varying the locations of the endpoint sites $s$ and $t$, retaining the best-scoring choice.

Following the procedure described in Section~\ref{subsec:workflow}, we first performed the simultaneous optimization of sequence and configuration space with $D=5$. After freezing the top-scoring geometry, we performed the sequence optimization with the full set of $D=20$ natural amino acids. The best-scoring element was then passed to ADCP, generating 100 ranked poses of this sequence in a off-lattice, all-atom representation.

\begin{figure*}
    \centering
    \includegraphics[trim=350 20 000 240, clip, width=1.23\textwidth]{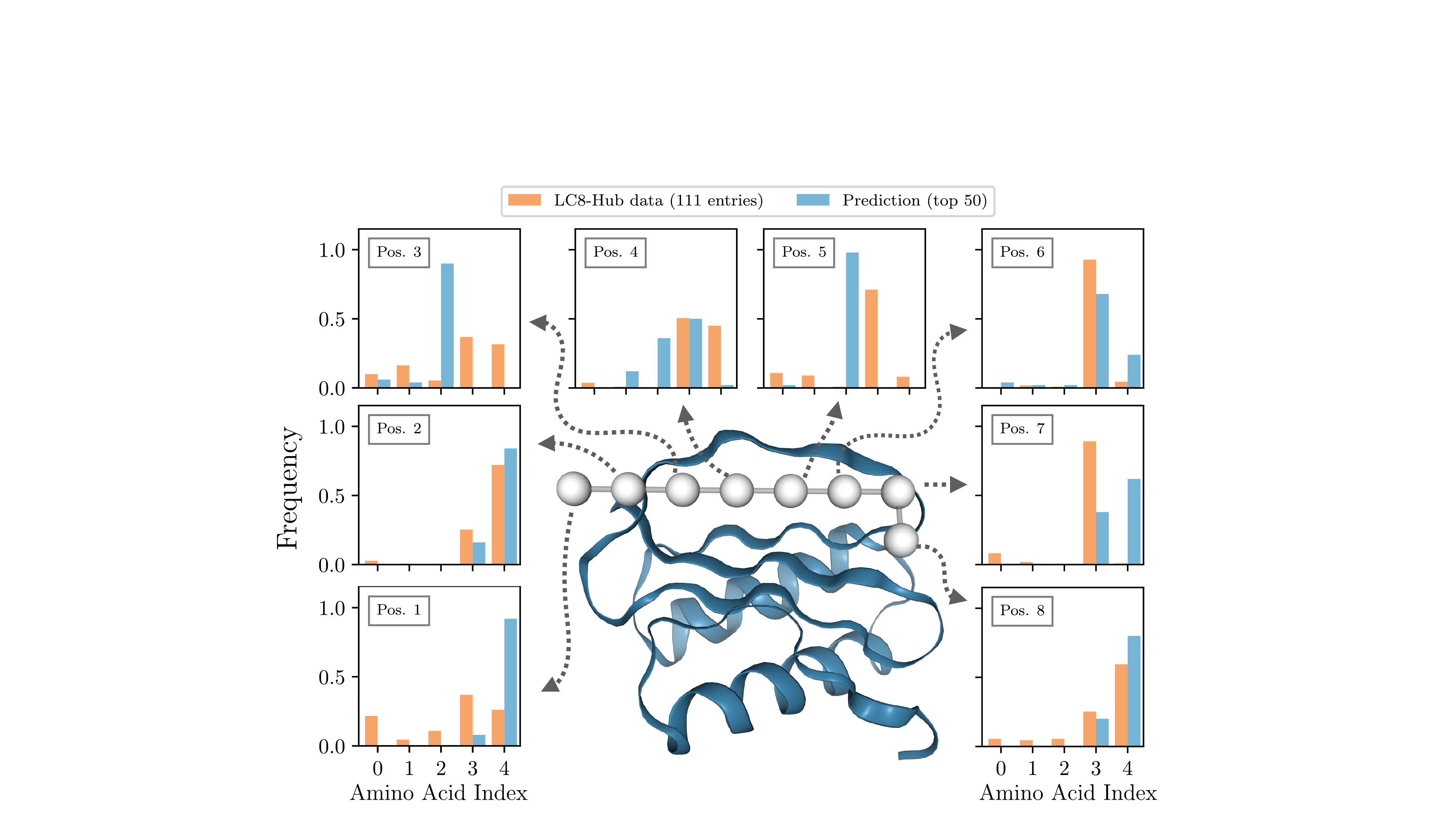}
    \caption{
    Comparison of the top 50 generated binders with 111 experimentally known binders of the LC8 hub protein. The eight histograms are corresponding to the eight positions in the anchor motif of LC8. For each position, the frequency of amino acid types in the generated binders (blue) and in the LC8 hub~\cite{LC8_binding_proteins_database} (orange) are displayed. The amino acids have been clustered into five groups as described in Section \ref{Section:CG_model}.
}
    \label{fig:histogram_LC8}
\end{figure*}
In general, the quality of these predictions depends on three main factors: (i) the accuracy of our coarse-grained energy model, (ii) the efficiency of the quantum optimization algorithm in identifying high-affinity sequences for the given pocket, and (iii) the reliability of the docking software in predicting the correct off-lattice binding pose.
To disentangle these factors, we used ADCP to perform two additional sets of calculations:\begin{itemize}
    \item Predicting binding poses for 30 randomly generated sequences.
    \item Redocking the peptide present in the experimentally resolved protein-peptide complex. 
\end{itemize}

The results of our precision-recall analysis for all three protein-peptide complexes are reported in Fig.~\ref{fig:histogram_AUC_PR_all}.
As expected, the redocking of the original peptide yields a high Area Under the Curve (AUC) score (orange vertical line) in all three cases. Conversely, the results based on docking randomly generated peptides are broadly distributed, with an average AUC score close to 0.4. 
The AUC of the peptides designed with our algorithm (blue vertical lines in Fig.~\ref{fig:histogram_AUC_PR_all}) are significantly closer to the experimental pose than the average AUC of the randomly generated peptides.
In particular, for protein 3BRL, the design algorithm yields remarkably good results, close to the ideal limit of our algorithm, set by the redocking curve.

We recall that our algorithm accounts for the peptide's average interactions using a mean-field approximation (Eq.~(\ref{eq:approx_sequence_FE})), enforcing the selectivity of the designed peptide to the given target. To investigate how our results are affected by this condition, we performed additional peptide design runs in which we neglected this factor. We found significantly worse results, as shown by the black vertical lines.
The fact that the resulting sequences perform even worse than the randomly generated ones suggests that neglecting $\langle U(\Sigma)\rangle_0$ introduces a systematic error. This effect is explained by the coarse-grained interaction energy being most attractive between hydrophobic residue pairs. Hence, optimizing the sequence to minimize only $U(\Gamma, \Sigma;P)$  yields hydrophobic sequences not specifically designed to match the chemical environment provided by the pocket.

\subsection{Sequence-Based Validation}\label{subsec:Sequence_Validation}

For a sequence-based validation of our design algorithm, we resorted to a dataset containing 111 peptides that bind to a specific pocket of the LC8 protein~\cite{LC8_Recognition_motifs,LC8_binding_proteins_database}, a molecular hub protein, which takes part in cell homeostasis.

As discussed in detail in~\cite{LC8_Recognition_motifs,LC8_binding_proteins_database}, the peptides in the database interact with the LC8 pocket via an 8-amino acid recognition motif.
We used our algorithm to design 50 different 8-residue peptides predicted to bind to the same pocket.
Our goal was to compare the primary sequences of the designed peptides with those in the experimental database.

\begin{figure}[t!]
    \centering
    \includegraphics[trim=5 0 0 0, clip, width=1\linewidth]{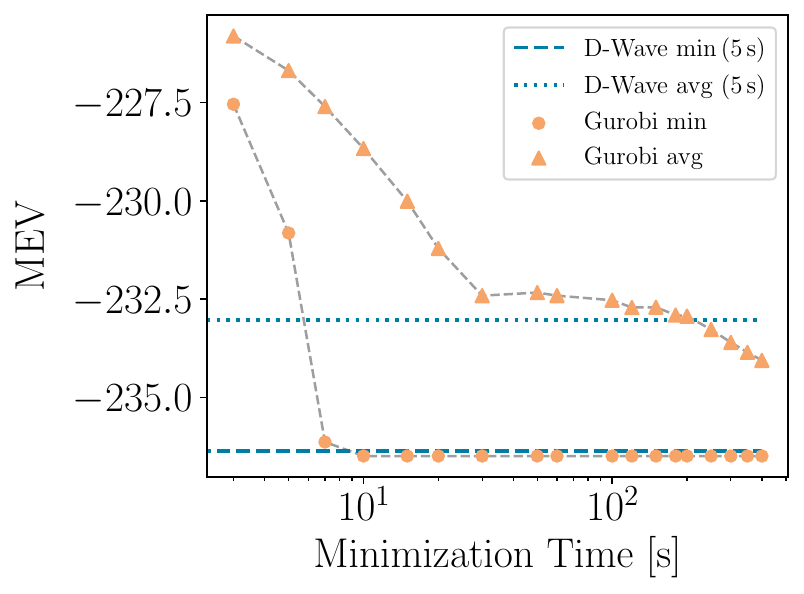}
    \caption{
    MEV obtained generating binders for the PDB entry 3BRL using classical and quantum optimization.
    Orange points: Lowest (circles) and average (triangles) MEV obtained in 1000 Gurobi runs as a function of the runtime of an individual simulation. Horizontal blue lines: lowest (dashed) and average (dotted) MEV obtained in 300 D-Wave runs with $5\,\text{s}$ of hybrid annealing time. 
    For a fair comparison, when running classical minimization with Gurobi, we encoded the conditions imposed by $H_{\text{anc}}$, $H_{\text{occ}}$, and $H_{\text{path}}$ as hard constraints.}
    \label{fig:DWave_Gurobi_MEV}
\end{figure}
As already mentioned, the chemical space of the designed peptide chains is huge, so it is unlikely that any design algorithm will generate sequences present in the dataset.
In Section \ref{Section:CG_model}, we exploited the redundancy of the amino acid alphabet to develop a coarse-grained model in which the 20 amino acid types were grouped into $ 5\lesssim D \lesssim 10$ families. 
The same procedure enables us to compare the designed and experimentally available sequences: We aim to identify correlations between the amino acid families found at different positions. 
In particular, the analysis reported in Fig.~\ref{fig:histogram_LC8} was performed by grouping the amino acids into 5 families.
Each of the eight histograms corresponds to one position in the binding motif, while the bars represent the relative frequency of the members of each of the 5 families: the blue (orange) bars refer to the relative population in the designed peptides (experimental dataset).
In comparing these two distributions, one should keep in mind that the experimental dataset may not provide an unbiased sample and might not represent the most optimal molecules for binding to the protein pocket. In spite of these limitations, the comparison between the distributions can still provide at least a qualitative assessment.
Overall, we find a good correlation between the histograms corresponding to the different datasets.
For example, at 6 out of the 8 positions (namely 1, 2, 4, 6, 7, and 8), the most frequently occurring amino acid family in the experimental dataset is among the two most frequently occurring families in the designed dataset. 
Positions 2, 6, and 8 show particular correlations between the datasets, suggesting that our design code can recognize which amino acid type is required at this position to promote binding.

To check that the observed correlation between the predicted and the observed sequences was not biased by the specific choice of the reduced alphabet size, in Fig.~\ref{fig:supp_lc8_10_clusters} of the SM, we report the results of an analog analysis in which the amino acids were grouped into $D=10$ families. As in the previous case, we observe an overall positive qualitative correlation. Similar to the results for 5 families, at positions 1, 2, 7, and 8, the amino acid family that is most frequently occurring in the experimental dataset is among the two most frequently predicted.
\subsection{Classical and Quantum Optimization}

The QUBO encoding introduced in Section~\ref{Section:quantum_encoding} enables us to resort to a classical optimizer or a quantum annealing machine to solve the optimization steps of our peptide design algorithm.
A key question to address is whether, for this specific problem, existing quantum annealing machines can compete with an industry-grade optimizer on a classical computer. 
To address this issue, we designed 10-amino-acid-long peptides for a protein-peptide complex investigated in Section \ref{subsec:Structure_Validation} (PDB entry: 3BRL) using D-Wave's hybrid classical/quantum solver (with default parameters) and Gurobi's classical optimization algorithm~\cite{gurobi}.
Our prescription to place the grid in the pocket (see Section \ref{subsec:Structure_Validation}) leads to lattices with typical sizes of $L_x$, $L_y\sim3$, and $L_z\sim10$ (see Section \ref{Section:quantum_encoding}). 
In our application to 3BRL, we used 1814 binary variables, in accordance with our estimate from Eq. (\ref{eq:qubit_estimate}).

To compare the quality and efficiency of the classical and quantum optimization algorithms, in Fig.~\ref{fig:DWave_Gurobi_MEV}, we report both the lowest and the average minimum-energy-value (MEV) attained by classical and quantum optimization. The results of Gurobi (version 11.0.1) were obtained by performing 1000 independent runs lasting between 3 and 400\,s. D-Wave's results were obtained with 300 independent 5\,s runs of the hybrid classical-quantum solver (hybrid\_binary\_quadratic\_model\_version2, with the quantum part executed on the performance-updated Advantage\_system).
These curves show that the two approaches yield a very similar lowest MEV, and that Gurobi stops improving on the lowest MEV after running for about 10\,s. 
On the other hand, to yield an average MEV lower than that generated by D-Wave, our Gurobi simulations need to run for approximately 200~s.

\begin{figure*}[t!]
    \centering
\includegraphics[trim=0 300 0 310, clip, width=0.95\textwidth]{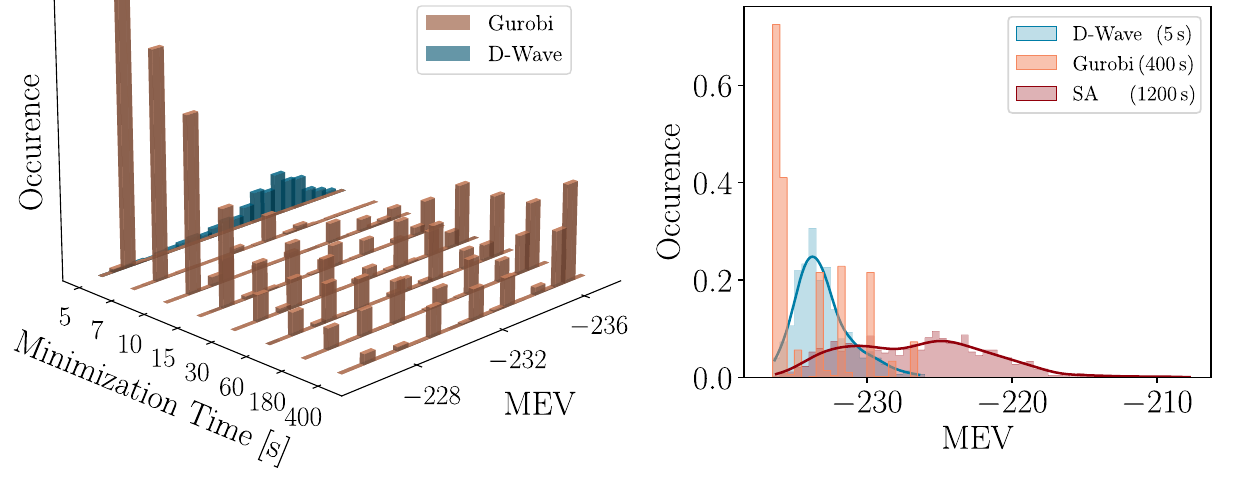}
    \caption{Spectrum of MEV obtained in classical and quantum optimization. Left panel: spectra obtained in 1000 Gurobi runs with different minimization times (brown) and in 300 5-s-long runs of the D-Wave hybrid. Right panel: spectra obtained with Gurobi run, with classical Simulated Annealing (SA) and with D-Wave's hybrid solver. }
    \label{fig:DWave_Gurobi_SA}
\end{figure*}

Designing a chemically diverse set of hits is important for efficient drug development. In our approach, this is obtained if the optimization does not yield a single MEV rather a distribution of diverse results peaked around a low average MEV.
In the left panel of Fig. \ref{fig:DWave_Gurobi_SA}, we compare the distributions of MEVs obtained using D-Wave and Gurobi.
We note that the quantum annealer yields a continuous energy spectrum of MEVs, while the distributions generated with Gurobi are peaked in a few isolated bins. By direct inspection, we found that each of such bins contains a single designed sequence.
Interestingly, running Gurobi longer does not lead to discovering more sequences. Instead, it only enhances the relative occurrence of the low-MEV sequences already discovered in shorter runs.
Simulated annealing represents a conventional classical optimization scheme that may be more apt to generate a continuum spectrum of MEVs. In the right panel of \ref{fig:DWave_Gurobi_SA}, we compare the spectra of MEVs generated with D-Wave and Gurobi with the results of 1000 simulated annealing independent runs, each consisting of $10^7$ sweeps and lasting approximately 1200\,s on a desktop computer.
As expected, the MEVs obtained by these relatively long simulated annealing runs are distributed according to a continuum spectrum. However, their average MEV is significantly higher than that obtained with quantum annealing runs lasting just 5\,s.
Collectively, this spectral analysis suggests that quantum optimization provides a promising approach to combine the request of high binding affinity (i.e., a low MEV) and chemical diversity.

These results show that, despite the outstanding limitations of the existing technology, quantum annealing machines can already be applied for realistic peptide design applications, yielding performances in line with those of a state-of-the-art classical optimizer. 
However, it should be emphasized that a quantitative relative assessment of these performances is not straightforward and goes far beyond the scope of the present work. Indeed, on the one hand, the efficiency of the classical optimization may be improved by resorting to more powerful computing resources. On the other hand, the quantum annealer's performance may be improved by tuning the internal parameters of the hybrid solver, such as the annealing time and schedule.

\section{Conclusions}\label{Section:Discussion}
The rapid development of quantum computing hardware raises the question of whether this emerging technology could accelerate computer-aided drug discovery. 
Pioneering applications of quantum algorithms to tackle drug discovery-related tasks include algorithms for molecular docking~\cite{docking_banchi_molecular_2020,docking_li_quantum_2024, docking_pandey_multibody_2022,docking_triuzzi_molecular_2024}, solvent configuration prediction~\cite{darcangelo_leveraging_2024}, sampling rare conformational protein transitions~\cite{ghamari_sampling_2024}, protein folding~\cite{folding_irback_2022, folding_babej_coarse-grained_2018, folding_conde-torres_classical_2024, folding_p_approach_2023, folding_robert_resource-efficient_2021}, and protein design~\cite{inv_folding_mulligan_designing_2019,inv_folding_irback_using_2024,inv_foldingpanizza_protein_2024}. Recently, Vakili \emph{et al.} combined classical and quantum neural networks to identify small molecules that inhibit KRAS proteins~\cite{ghazi_vakili_quantum-computing-enhanced_2025}.

In this work, we developed a physics-based multi-scale approach to \emph{de novo} peptide design that exploits the potential of quantum hardware to enhance the simultaneous exploration of all possible peptide's sequences and conformational states.
We derived its quantum encoding starting from a rigorous statistical mechanical formulation -- condition~(\ref{exactCondition}) -- by applying a leading-order cumulant expansion -- condition~(\ref{cumulant})-- and a mean-field approximation -- Eq.~(\ref{eq:approx_sequence_FE})--. 
Our scheme resorts also to classical computing to improve the structural resolution of the binding pose and yield atomically resolved off-lattice predictions.

We illustrated this approach with several applications, comparing our results against experimentally characterized protein-peptide complexes.
In a first validation study, we assessed the reliability of our structural predictions by comparing the binding poses of designed peptides to those obtained by redocking the peptides present in the corresponding experimentally resolved complexes. 
In a second validation study, we statistically compared the sequences generated by our algorithm to bind to the protein LC8 with those present in a dataset of experimentally verified peptide binders.
Both validation studies suggest that our algorithm successfully generates molecules with the desired structural and chemical properties.

In future work, it will be important to perform direct experiments aiming at assessing the binding affinity of the predicted sequences to the target protein. Other relevant improvements would be to generalize our peptide binder design scheme to small molecules and to account for receptor flexibility and ADMET properties in the optimization process. 

We compared the solutions to our design problem obtained using the D-Wave quantum annealer with those obtained using conventional simulated annealing and Gurobi, an industry-grade classical optimizer. Even using modest qubit resources, D-Wave in a few seconds generated sequences with MEVs close to those obtained with Gurobi on a desktop computer, and significantly lower than those generated with much longer simulated annealing runs. 
We found that quantum optimization yields a continuum spectrum of MEVs, while Gurobi tends to systematically converge towards a discrete set of MEV solutions. Direct inspection revealed that all 300 MEVs obtained with D-Wave correspond to different peptide sequences. Therefore, the results obtained by quantum optimization combine a good predicted affinity with chemical diversity. Overall, our results show that, even in their current early stage of development, quantum computers can already be useful for physics-based drug design.

With increasing system sizes and with more detailed molecular representations, we expect the energy landscape generated by the QUBO Hamiltonian resemble that of a spin glass, leading to an exponential growth of complexity of the related discrete optimization problem. 
Our classical simulations were conducted on a desktop computer and may be sped up using more considerable computing resources. However, classical algorithms and hardware are already highly optimized, and state-of-the-art solvers like Gurobi do not profit significantly from GPU acceleration. In contrast, quantum technologies are still in their infancy. If the hardware continues to improve over the next several years, quantum optimization algorithms may enable tackling complex \emph{de novo} drug design problems out of the reach for classical machines. 

Furthermore, similar quantum-empowered physics-based design approaches may be developed for applications beyond drug discovery, such as the design of organic semiconductors or molecular nanosensors.
\newline
\newline
{\emph {\bf Acknowledgments}: We are grateful to V. Panizza and C. Micheletti for useful discussions. }
\bibliography{Citations}

\clearpage
\appendix
\renewcommand{\thefigure}{S\arabic{figure}}
\renewcommand{\thetable}{S\arabic{table}}
\renewcommand{\theequation}{S\arabic{equation}}
\renewcommand{\thesection}{S\arabic{section}}
\setcounter{figure}{0}
\setcounter{table}{0}
\setcounter{equation}{0}
\setcounter{section}{0}
\renewcommand{\appendixname}{}
\section*{Supplementary Material}
This Supplementary Material details the implementation of our design scheme, its quantum encoding, and the statistical analysis we performed to assess its reliability. 

In particular, Section~\ref{SM:AverageContacts} describes how we computed the number of contacts per amino acid, a key ingredient needed in the mean-field estimate of the $\langle U_T(\Sigma)\rangle$ term of Eq.~(\ref{cumulant}) of the main text. Section~\ref{SM:Scaling} reports an estimate of the number of qubits needed to generate peptides with our algorithm. 
Sections~\ref{SM:docking_validation} and~\ref{SM:LC8_more_clusters} report details and additional results of our validation studies.
 
\section{Estimate of the Number of Contacts per Amino Acid}\label{SM:AverageContacts}
In Eq.~(\ref{eq:approx_sequence_FE}) of the main text,  we estimate the average interaction of the peptide with the protein surface. In particular, for each amino acid in the sequence, we calculate the interaction it would form with any other amino acid on the protein surface according to the relative frequency it occurs on surfaces. The last ingredient needed to calculate the average interaction is the average number of contacts $\mathcal{N}_c$ formed by the sequence with the pocket.

According to the LJ potential introduced in Eq.~(\ref{eq:LJ_potential}) of the main text, the maximal interaction between a pair of amino acids $i$ and $j$ is $\varepsilon_{ij}$. 
We therefore define a partial contact as the fraction of the maximal interaction it forms, i.e., we obtain the partial contacts associated with a given pairwise interaction by dividing the interaction energy by $\varepsilon_{ij}$.
Summing over all partial contacts and dividing by the number of amino acids in the ligand yields the average number of contacts $\mathcal{N}_c$.

We note that the number of contacts is a specific property of a given model for the amino acid interaction. For example, increasing the cutoff of the LJ potential leads to a larger $\mathcal{N}_c$.

In practice, we first performed a simulation with $\mathcal{N}_c=0$. The number of contacts of the resulting binding pose was then used as the input to a new simulation. If the number of contacts was significantly different (roughly by 10\,\%), the procedure was repeated, initializing a new simulation with the updated value.
\section{Scaling of the Number of Qubits}\label{SM:Scaling}
In the following, we give an estimate on how many binary variables (or qubits) are needed to run our peptide design algorithm.
Let $N_{\text{Lattice}}$ be the number of lattice points and $N_{\text{Degree}}$ be the degree of the lattice points, i.e., the number of bonds that can be assigned to each lattice point. For simplicity, we assume that the degree is the same for each grid point. As a bond connects two lattice points the total number of bonds is $N_\text{Bonds,tot}=N_{\text{Lattice}}N_{\text{Degree}}$/2. Furthermore let $D$ be the number of chemical building blocks per grid point. A set of $D$ qubits $\{q^{(k)}_{i}\}_{k\in\set{1, \dots, D}}$ is associated to each lattice point $i$. This amounts to $D_\text{tot}=DN_{\text{Lattice}}$ binary variables, representing all building blocks on all grid points. 
Furthermore, for the implementation of nearest neighbour interactions on a quantum annealer, a set of ancillary variables
$\{ q^{(k)}_{ij}\}^{k\in\set{1,\dots,D}} _{ij \in \set{1,N_{\text{Bonds, tot}}}}$ is required which amounts to $N_{\text{Anc}}=DN_{\text{Bonds, tot}}$ extra variables.
Note that for next-nearest neighbour interactions, no additional variables are needed.
The total number of variables is therefore 
\begin{align}
N_\text{tot} &=
D_\text{tot}+N_{\text{Bonds, tot}} + N_{\text{Anc}} \notag \\
    &=N_{\text{Lattice}}\left( D+\frac{N_{\text{Degree}} +DN_{\text{Degree}}}{2}\right)
\end{align}
In the case of a cubic lattice of dimensions $L_x, L_y, L_z$ the precise resource requirements amount to
\begin{align}
    D_\text{tot}&=DL_xL_yL_z\notag \\
    N_{\text{Bonds, tot}} &= (L_x-1)L_yL_z+L_x(L_y-1)L_z+L_xL_y(L_z-1) \notag \\
    N_{\text{Anc}} &= N_{\text{Bonds, tot}}D \notag \\
    N_{\text{tot}} &= N_{\text{Bonds, tot}}(D+1) +D L_xL_yL_z
\end{align}
In our implementation, the choice of hardware was D-Wave's quantum annealing platform. However, we could have also opted for gate-based quantum computers, optimizing the energy using, e.g., the quantum approximate optimization algorithm~\cite{farhi_quantum_2014} or variational quantum eigensolvers~\cite{peruzzo_variational_2014}.
Although currently available quantum annealers have more qubits than gate-based quantum computers, implementing the algorithm on a gate-based quantum computer requires fewer qubits.
In particular, the ancillary qubits implementing the nearest neighbour interactions are not required on a gate-based quantum computer.
We could further reduce the number of lattice qubits using a logarithmic encoding from $N_{\text{Lattice}}$ to $\log(N_{\text{Lattice}}+1)$.
We would therefore require 
\begin{align}
    N_{\text{BB, tot}}&=L_xL_yL_z\log_2(D+1) \notag \\
    N_{\text{Bonds, tot}}&=(L_x-1)L_yL_z+L_x(L_y-1)L_z+L_xL_y(L_z-1) \notag\\
    N_{\text{tot}} &= N_{\text{Bonds, tot}} + \log_2(D+1)L_xL_yL_z
\end{align}
qubits on a gate-based quantum computer.
However, the number of qubits is not the only relevant measure in terms of resource requirements. For example, a logarithmic encoding introduces higher-order interaction terms, making the implementation of the Hamiltonian more involved.

\section{Details on the Structure-Based Validation}\label{SM:docking_validation}
In Section~\ref{subsec:Structure_Validation} of the main text, we described a structural validation of design algorithm. 

Note that this analysis was performed by resorting to local docking. This might explain the high scores associated to some of the randomly generated peptides.
However, in at least one of the protein-peptide complexes we considered, the global docking did not yield good results, even for the redocking experiment so could not be used to validate our algorithm. 

Using the recommended simulation parameters (3 million steps per amino acid and 200 independent runs) each local redocking calculation took roughly 3 hours on a desktop computer.

\section{Additional Results on the Sequence-Based Validation}\label{SM:LC8_more_clusters}
\begin{figure*}
    \centering
    \includegraphics[width=1.0\linewidth]{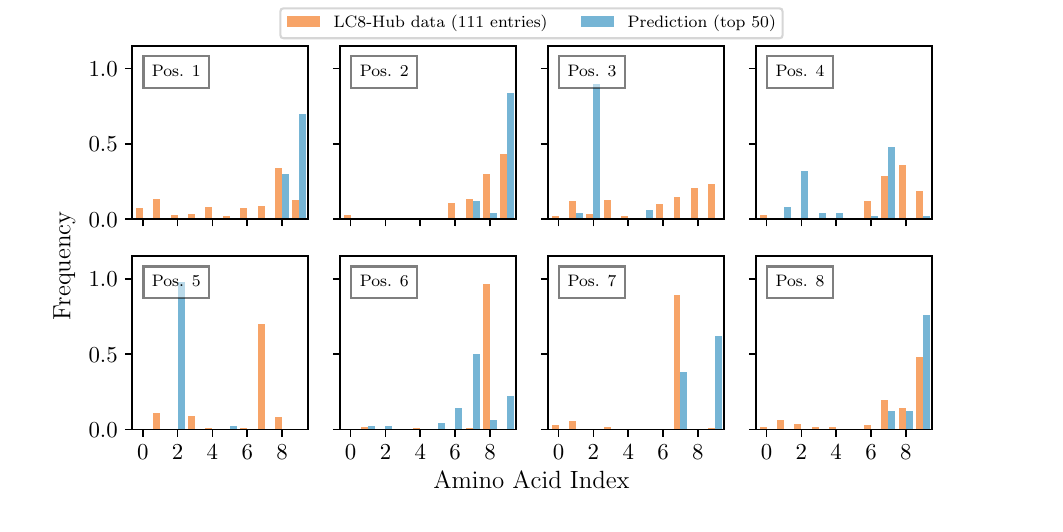}
    \caption{Comparison of proposed generated binders and known binders of the LC8 hub protein as in Fig.~\ref{fig:histogram_LC8}. The amino acids have been clustered into 10 amino acid types.}
    \label{fig:supp_lc8_10_clusters}
\end{figure*}

In Section~\ref{subsec:Sequence_Validation} of the main text, we generated ligands for the LC8 hub protein and compared their sequences to the sequences of known binders.
We clustered the amino acids into five families of similar types before comparing the relative occurrences at each position of the LC8 binding motif, as shown in Fig.~\ref{fig:histogram_LC8} of the main text.
In Fig.~\ref{fig:supp_lc8_10_clusters},  we present similar results obtained by clustering the amino acids into 10 families.
Even at this finer resolution, we still find that in four out of eight positions, namely, in positions 1, 2, 7 and 8, the most frequently experimentally occurring amino acid type is among the two most frequently predicted ones. In position 4, the most frequently predicted amino acid type is the second most occurring one.

\end{document}